\documentclass[useAMS,usenatbib]{mn2e}

\usepackage[T1]{fontenc}
\usepackage{longtable}
\usepackage{color}
\usepackage{array}
\usepackage{graphicx}
\usepackage{caption}
\usepackage{url}
\usepackage{aecompl}

\title[Pal 5 and its Tidal Tails]{Palomar 5 and its Tidal Tails: A Search for New Members in the Tidal Stream}
\author[P. B. Kuzma et al.] {P. B.~Kuzma,$^{1}$, G. S.~Da~Costa$^{1}$, S. C.~Keller$^{1}$, E.~Maunder$^{1}$\\
$^{1}$Research School of Astronomy \& Astrophysics, Australian National University, Mount Stromlo Observatory\\ via Cotter Road, Weston, ACT 2611, Australia; pete.kuzma@anu.edu.au\\ }

\begin{document}

\date{}

\pagerange{\pageref{firstpage}--\pageref{lastpage}} \pubyear{2002}

\maketitle

\label{firstpage}

\begin{abstract}

In this paper we present the results of a search for members of the globular cluster Palomar 5 and its associated tidal tails. The analysis has been performed using intermediate and low resolution spectroscopy with the AAOmega spectrograph on the Anglo-Australian Telescope. Based on kinematics, line strength and photometric information, we identify 39 new red giant branch stars along $\sim$20$^{\circ}$ of the tails, a larger angular extent than has been previously studied. We also recover eight previously known tidal tail members.  Within the cluster, we find seven new red giant and one blue horizontal branch members and confirm a further twelve known red giant members. In total, we provide velocity data for 67 stars in the cluster and the tidal tails. Using a maximum likelihood technique, we derive a radial velocity for Pal 5 of $-57.4 \pm 0.3$ km s$^{-1}$ and a velocity dispersion of $1.2\pm0.3$ km s$^{-1}$. We confirm and extend the linear velocity gradient along the tails of $1.0 \pm 0.1$ km s$^{-1}$ deg$^{-1}$, with an associated intrinsic velocity dispersion of $2.1\pm0.4$ km s$^{-1}$.  Neither the velocity gradient nor the dispersion change in any significant way with angular distance from the cluster, although there is some indication that the gradient may be smaller at greater angular distances in the trailing tail.  Our results verify the tails as kinematically cold structures and will allow further constraints to be placed on the orbit of Pal 5, ultimately permitting a greater understanding of the shape and extent of the Galaxy's dark matter halo. 
\end{abstract}


\section{Introduction}

The globular clusters (GCs) of the Milky Way have proven to be treasure chests of invaluable information about the Galactic halo. These stellar systems are self-gravitating groups of similar stars, in both age and metallicity. Stars that reside in the outer regions of a cluster are sensitive to the gravitational tidal field of the Galaxy, and if the cluster potential is overcome, the stars can be lost from the cluster to the halo field.  The distance from the cluster centre at which the gravitational forces are balanced is known as the tidal radius, $r_{t}$. The fitting-formulae of \cite{1962AJ.....67..471K} and more sophisticated modelling \citep[e.g.,][]{2005ApJS..161..304M} have provided reasonable estimates of the tidal radius for a number of GCs. However, some clusters do not exhibit a classic tidally-limited profile, revealing instead ``extra-tidal'' features.  For example, \cite{1995AJ....109.2553G} showed through star counting techniques the existence of clusters that have density profiles extending well beyond the limiting radii set by the best-fit King profile. These extra-tidal features are generally indicators of a significant loss of stars from the cluster as a result of tidal interactions with the Galaxy, potentially leading to the complete disruption of the cluster.   The escaping stars form leading and trailing streams (tidal tails) generally aligned with the orbit of the cluster. Consequently, the tidal tails present a prime opportunity to further define the orbit of the parent cluster, which in turn allows constraints to be placed on the potential field of the Galaxy's dark matter halo.

Amongst the Galactic globular clusters with extra tidal features, Palomar 5 (Pal 5) stands out. At a distance of 23.2 kpc from the Sun \citep[e.g.,][]{2012A&A...546L...7M}, the cluster has a number of characteristics (low luminosity, low central concentration and low velocity dispersion) that made it a perfect candidate for a cluster undergoing tidal disruption. \cite{2001ApJ...548L.165O} uncovered the presence of substantial tidal tails through spatial analysis techniques utilizing the extensive photometry provided by the Sloan Digital Sky Survey (SDSS; \cite{1998AJ....116.3040G}; \cite{1996AJ....111.1748F}; \cite{2000AJ....120.1579Y}; \cite{2009AJ....137.4377Y}). Later data releases of the SDSS allowed additional analysis: \cite{2006ApJ...641L..37G}, following similar techniques to \cite{2001ApJ...548L.165O}, extended the definition of the trailing tail to roughly 16$^{\circ}$ from the cluster centre and that of the leading tail to $\sim$6$^{\circ}$, at which point the SDSS coverage ends. With further analysis, the trailing tail has now been shown to span at least 23$^{\circ}$ from the cluster centre, where again the limits of the SDSS survey area are reached \citep{2012ApJ...760...75C}.

The discovery of the tails spurred further study of Pal 5. For example, \cite{2004AJ....128.2274K} noted that the luminosity function (LF) of Pal 5 in the cluster core is flatter than the LF in the outer regions. This indicates that the core of Pal 5 lacks low mass stars as result of dynamically driven mass segregation. \cite{2004AJ....128.2274K} also investigated the LF of the tails, noting that it is comparable to that for the outer regions of cluster. These results complement those of \cite{2003AJ....126.2385O} who report that the mass in the tails is greater than the mass remaining within the cluster. It is likely that another passage of Pal 5 through the disk of the Galaxy will prove to be the final one before the cluster is completely disrupted \citep{2003AJ....126.2385O}. 

In this respect \cite{2004AJ....127.2753D} completed a large number of N-body simulations of clusters travelling along an orbit analogous to that of Pal 5 in the potential of the Milky Way. The simulations showed that clusters with similar properties to Pal 5 would create tidal tails from multiple passages through the disk of the Galaxy, and that these disk crossings can eventually lead to the complete dissolution of the cluster. Indeed these simulations predict the complete destruction of Pal 5 at its next disk crossing. 

Nonetheless, the simulations failed to produce some of the structure seen within the Pal 5 tails. In particular, as first noted by \cite{2003AJ....126.2385O}, the tails display a series of inhomogeneities along their length, visible as regions of higher and lower density \citep[see also][]{2006ApJ...641L..37G, 2010A&A...522A..71J}.  \cite{2012ApJ...760...75C} suggested that the inhomogeneities may have been created by the interaction of the stream with dark matter sub-halos present in the Galactic halo, potentially providing an important probe of the predictions of the standard $\Lambda$CDM model for the Galaxy (\cite{2014ApJ...788..181N} and references therein). However, \cite{2012A&A...546L...7M} showed through detailed $N$-body simulations that the clumps and gaps can also result from the epicyclic motion of the stars in the tidal tails.  The question has been further investigated with the simulations performed by \cite{2014ApJ...788..181N}.   These showed that in a $\Lambda$CDM Milky Way dark matter halo model, gaps in tidal streams can be caused by both purely epicyclic motions and by sub-halo interactions, with the presently available data unable to definitely distinguish between the possibilities. Most recently,  \citet{2014arXiv1410.3477P} found that the thin shape of the tails can be successfully reproduced in  spherical dark matter halo potentials. However, they found this is not the case for the triaxial potential proposed by \cite{2010ApJ...714..229L} to describe the properties of the Sagittarius stream.

The kinematics of the cluster itself have been studied by \cite{2002AJ} (hereafter O02). O02 found the heliocentric velocity of the cluster to be $-58.7 \pm 0.2$ km s$^{-1}$ with a notably small velocity dispersion of $1.1 \pm 0.4$ km s$^{-1}$. Subsequently, \cite{2009AJ....137.3378O} (O09) provided a kinematic analysis of individual stars in the tails of Pal 5. Seventeen stars were determined to be members of the tails based on their line-of-sight velocities. As for the cluster the tails were shown to have a low velocity dispersion: $\sigma$ < 5 km s$^{-1}$. Such a low dispersion is a defining characteristic of a kinematically cold structure. The velocities of the stars along the tails also revealed a velocity gradient of $\sim$1 km s$^{-1}$ deg$^{-1}$. These results suggested a revision of the orbit of Pal 5, and O09 further found that the results are best interpreted if the tails do not align exactly with the orbit of Pal 5, contrary to earlier indications \citep{2001ApJ...548L.165O}. O09 point out the  need for additional kinematic information at larger distances along the tail to further constrain the simulations of the orbit. \cite{2013MNRAS.436.2386L} reach similar conclusions.

In this paper we present a self-consistent analysis to identify additional members of Pal 5 and of its tidal tails.  In particular we explore the  full $20^{\circ}$ extent of the tails presented in \cite{2006ApJ...641L..37G}.  In the following section we describe the observations and the analysis techniques employed.  In section 3 we discuss our results, first for the cluster and then for the stars in the tidal tails.  Section 4 contains our concluding comments. 

\begin{figure*}
  \begin{center}  
     \includegraphics{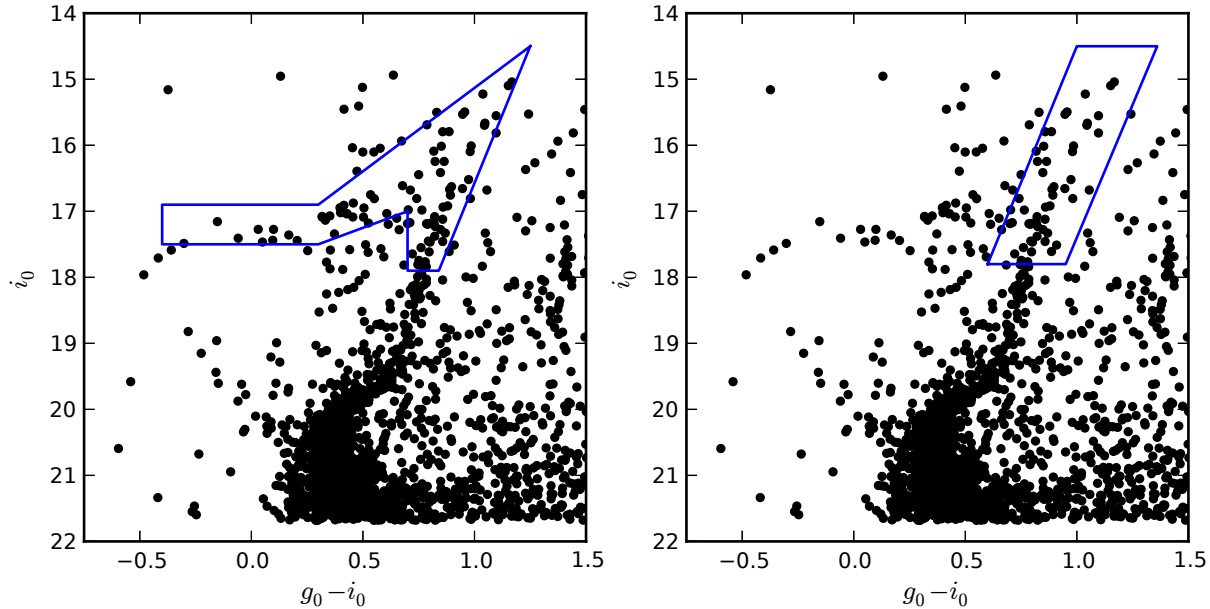}
     \end{center}
   \caption{Both panels show a dereddened colour-magnitude diagram for Pal 5 using photometry from SDSS DR10 \citep{2014ApJS..211...17A}. Only stars within 8.3$^{\prime}$ of the cluster centre are plotted.  In the left panel the blue polygon outlines the approximate target selection region for the observations conducted in 2006, 2009 March and 2010 May.  The right panel shows the target selection region for the 2009 April observations. }
\label{fig:p5cmd}
\end{figure*}

\begin{table*}
\centering
\begin{minipage}{170mm}
\caption{List of observations. Fields have been named 1 -11 based on increasing R.A.}
\label{tab:obslist}
\begin{tabular}{@{}ccccccccc@{}}

\hline \hline
Field Name& \multicolumn{2}{c}{Mean Field Center} & Date-obs &\# of config.&\# of exp.&Exp. Time &Red Grating&Blue Grating\\
 & R.A (J2000) & Dec (J2000)&mm/year&&&per Obs. (s)&&\\
\hline
F1&15:09:44.89&--01:48:00.1&04/2009&2&2&1200&1000I&580V\\
F2&15:10:00.57&--01:29:58.5&06/2006&1&3&1800&1700D&2500V\\
F3$^{*}$&15:15:59.93&--00:06:37.4&06/2006&3&3&1800&1700D&2500V\\
&&&03/2009&1&3&1200&1000I&580V\\
F4$^{*}$&15:17:59.67&00:19:59.6&06/2006&1&3&1800&1700D&2500V\\
&&&05/2010&1&3&1200&1700D&580V\\
F5&15:23:59.97&01:30:00.1&05/2010&1&4&1200&1700D&580V\\
F6&15:31:59.37&03:29:54.9&06/2006&1&3&1800&1700D&2500V\\
&&&03/2009&1&3&1200&1000I&580V\\
F7&15:40:00.17&04:00:01.3&06/2006&2&3&1800&1700D&2500V\\
F8&15:48:01.19&04:41:55.6&06/2006&1&3&1800&1700D&2500V\\
F9&15:56:00.82&05:30:03.3&06/2006&2&3&1800&1700D&2500V\\
F10&16:03:12.70&06:30:01.6&06/2006&2&3&1800&1700D&2500V\\
F11&16:16:49.07&07:47:55.2&06/2006&1&3&1800&1700D&2500V\\
\hline
 
\end{tabular}
$^{*}$ These fields contain the cluster centre.

\end{minipage}
\end{table*}

\section{Observations And Data Reduction}
\subsection{Observations and Target Selection}
The observations employed for this work were taken with the Anglo-Australian Telescope (AAT) at Siding Spring Observatory (SSO), using AAOmega, a multi-fibre, dual-beam spectrograph that utilizes the two degree Field (2dF) fibre-positioning system\footnote{Manuals and technical information at \url{http://www.aao.gov.au/2df/aaomega/}}. The system can allocate up to 392 fibres allowing simultaneous observations of both science targets and sky regions across a $2^{\circ}$ diameter field-of-view. The light fed into the spectrograph is split into the red and blue arms by a dichroic centred at 5700\AA. This work makes use of a number of observations performed across five years.  These include our own observations from 2009 and 2010, as well as a set from 2006 (PI: Lewis) obtained from the AAT archive. In the 2006 June observations, 14 2dF configurations were observed over five nights at nine distinct field centres spread along the leading and trailing tails.  The total integration time per configuration was 3$\times$30 min.  For this run the red arm of AAOmega was configured with the 1700D grating and the blue arm with the 2500V grating.  The red arm spectral coverage was  8450 -- 9000\AA\/ at a resolution of $\mathcal{R} \approx 10000$ while for the blue arm the coverage was 5280 -- 5630\AA\/ at $\mathcal{R} \approx 8000$. 

The second set of observations took place in 2009 March and April. Completed during service observing runs, AAOmega was configured with the 1000I grating (spectral range: 8000 -- 9500\AA\, with a coverage of 1100\AA, $\mathcal{R}=4400$) in the red arm and the 580V grating (spectral range: 3700 -- 5800\AA\, full coverage,  $\mathcal{R}=1300$) in the blue arm.  In 2009 March single configurations were observed at two field centres while in 2009 April two configurations were observed at a field centre located in the leading tail. The integration times were 3$\times$20 min for the March observations and 2$\times$20 min for the April set.  The final set of observations used for this work took place in 2010 May, with AAOmega configured with the 1700D (red arm) and 580V (blue arm) gratings. Single configurations at two field centres were observed with integration times of 3$\times$20 min and 4$\times$20 min, respectively.  Overall each 2dF configuration typically consisted of approximately 330 targets together with 30 fibres allocated to blank sky regions. Table \ref{tab:obslist} gives an overview of all the observations used in this work; the total number of stars observed was 4507.

The selection of stars targeted for observation with 2dF varied across the different runs, and this is illustrated in Fig. \ref{fig:p5cmd}.  Both panels display the reddening corrected colour-magnitude diagram (CMD) for Pal 5 generated from the SDSS DR10 photometry \cite{2014ApJS..211...17A}. Only stars within 8.3$^{\prime}$ of the cluster centre are plotted and the reddening corrections made use of the dust maps available from \cite{1998ApJ...500..525S}.  In the left panel the approximate region used for target selection for the 2006, 2009 March and 2010 May observations is delineated, while the right panel shows the approximate target selection region for the 2009 April run.  The reddening corrections to the SDSS
photometry were small, as there is little variation from the cluster value of E$(B-V)$ = 0.06 \citep{1998ApJ...500..525S} across the regions of the tidal tails studied.

\subsection{Reduction and Techniques}
Once the data had been extracted from the AAT archive, it was reduced using the 2dF data reduction pipeline, 2dfdr\footnote{Visit \url{http://www.aao.gov.au/2df/aaomega/aaomega_2dfdr.html} for more information}. The approach was the standard one using fibre flats to set the location of the spectra, and arc lamp spectra for the wavelength calibration.  The relative throughput of the fibres, necessary for the sky subtraction, was determined using the \textit{SKYFLUX(MED)} approach, which determines the relative throughputs from the observed intensities of night-sky emission lines.  At the end of the process the wavelength-calibrated sky-subtracted spectra from the individual integrations were median-combined to remove any cosmic-ray contamination.  Typical signal-to-noise ratios (S/N) range from 15 to 70 pixel$^{-1}$ in the vicinity of the Ca II triplet for the red spectra, and 10 to 40 pixel$^{-1}$ in the vicinity of the Mg~I lines in the blue-arm spectra.

\subsubsection{Radial Velocities}
The radial velocities of the stars were calculated from the red arm spectra via cross-correlation using the \textbf{IRAF}\footnote{Information and distribution of IRAF is available through \url{http://iraf.noao.edu/}.} routine \textit{fxcor}. The template used for the correlation was an
AAOmega 1700D grating, high signal-to-noise ($>$100) spectrum of the F6V star HD 160043 taken as part of the program
described in \cite{2012ApJ...751....6D}.  The strength of the Ca~II triplet lines in this star match well with those in the 
program object spectra.  The spectra were correlated over the wavelength interval 8450\AA$\,< \lambda <\, $8700\AA, a region relatively
uncontaminated by night-sky emission line residuals. Heliocentric velocities of the targets were calculated with the \textbf{IRAF} command \textit{rvcorrect}, and, as discussed in \cite{2012ApJ...751....6D}, the uncertainty in the zero point of the radial velocity system
is $\pm$0.8 km s$^{-1}$. Stars that had low correlation peak heights (<0.5) and/or high uncertainties in the correlation velocity (>5 km s$^{-1}$) were discarded from the subsequent analysis -- generally these were spectra with low signal-to-noise.

A number of stars were observed across multiple fields. We used these multiple observations to estimate the overall accuracy of the 
velocities returned by the \textit{fxcor} routine. The mean velocity of stars with two or more observations was calculated using the output errors of \textit{fxcor} as weights.  The corresponding estimate of the error for a single observation was then evaluated using the small number statistics formalism of \citep{Keeping:1995vg}, which utilizes the range of the observations.  In particular, the estimated standard deviation for a single observation is given by: 
\begin{equation}
\sigma=R\times q_{N} \label{eq:stdev}
\end{equation}
where $R$ is the range in $N$ observations and $q_{N}$ is a multiplicative factor (e.g., $q_{2}$ = 0.886 and $q_{3}$ = 0.591). 
We then compiled these error estimates as a function of the median signal in the continuum region between the stronger Ca II lines, finding that for stars with a median continuum level above 1200 ADU the single observation error estimate was less than 1 km s$^{-1}$. As the continuum level decreases, the velocity error increases towards 2 km s$^{-1}$ at continuum levels $\sim$700 ADU and then increases rapidly to $\sim$4 km s$^{-1}$ at $\sim$200 ADU\@.  These results are consistent with those of \cite{2012ApJ...751....6D} who used a similar instrumental setup and analysis technique.   We employed this ($\sigma_{v}$, continuum level) relation to generate the velocity uncertainty estimates for stars with only one observation.  For stars with multiple observations the estimate was reduced by the square-root of the number of observations.

\subsubsection{Photometric Discrimination}  \label{sect:2.2.2}
Although the primary targets were the stars in the selection boxes shown in Fig.\ \ref{fig:p5cmd}, the actual observations included stars with a broader range of colours so that as many of the available 2dF fibres were allocated as possible.  However, no unusual stars were discovered, and since there is no reason to expect any Pal 5 tidal tail stars to lie significantly away from the principle sequences in the CMD, in the subsequent analysis we focus only on those stars that lie relatively near to the Pal 5 sequences in the CMD\@. A routine was created to remove stars from the data set if their CMD location did not lie within a polygon encompassing the Pal 5 CMD features, similar to that shown in the left panel of Fig.\ \ref{fig:p5cmd}. 

\begin{figure*}
  \begin{center}  
    \includegraphics{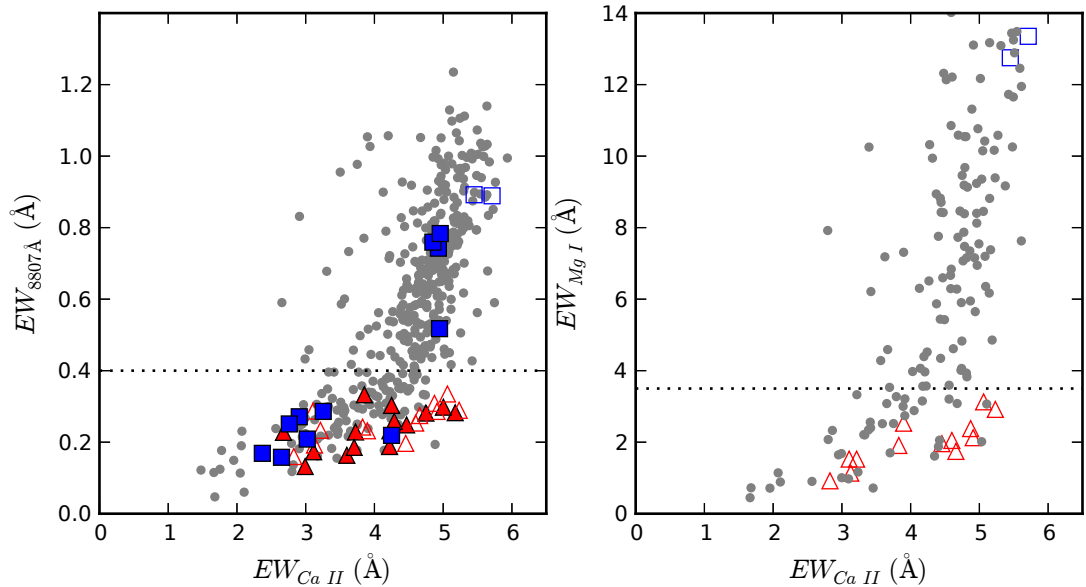}
  \end{center}
\caption{Left panel: Equivalent width (EW) of the $\lambda$Mg~I 8807\AA\/ line as function of the sum of the EWs of the Ca II triplet lines at $\lambda\lambda$8542 and 8662\AA\/ for stars in fields F3 and F4, which contain Pal 5.  Stars that lie beyond the adopted radius of the cluster (8.3$^{\prime}$) are plotted as grey points. Blue squares show stars within the adopted cluster radius but outside the velocity range $-65$ to $-50$ km s$^{-1}$, while red triangles show stars within the adopted radius and within the velocity range.  In both cases open symbols are used for stars that are also plotted in the right panel.  Right panel: EW of the Mg~I triplet features at $\lambda\lambda$5172\AA\/ plotted against the summed EW of the Ca II triplet lines, using the same symbols as for the left panel.  The dotted line in both panels shows the Mg I EW values used to separate dwarfs from giants.}
\label{fig:ewmgca}
\end{figure*}

\subsubsection{Giant/Dwarf discrimination}
The principle contaminant in the fields containing the Pal 5 cluster and tidal tail stars, which are giants, are foreground dwarfs of approximately solar metallicity.  A means of distinguishing these stars from the potential cluster and tidal tail members is therefore needed.  
We adopt a similar approach to that
of \citep{Battaglia2012}, which employs the gravity sensitivity of the Mg I line at $\lambda$8807\AA, a line which is stronger in dwarfs than in giants of similar temperature and metallicity.  The discrimination is aided by the fact that the Pal 5 giants are also metal-poor compared to the vast majority of field dwarfs.  In left panel of Fig. \ref{fig:ewmgca}, we show the relationship between the equivalent width (EW) of the Mg~I 
$\lambda$8807\AA\/ line and the sum of the EWs of the two stronger Ca II triplet lines at $\lambda\lambda$8542 and 8662\AA\/ for the stars in the
two fields which contain the cluster centre. The EW measurements were made using the routine \textit{splot} in \textbf{IRAF}. The uncertainties in the line strengths were estimated from the stars with multiple observations and are typically 0.15\AA\/ in size.  We also identify in the Figure stars that lie within our adopted radius for Pal 5 (8.3$^{\prime}$; see \S \ref{sect3}) and within the velocity range encompassing cluster members.  As expected, these probable giant stars occupy the lower part of the relationship.  We therefore classify as dwarfs those stars with Mg~I $\lambda$8807\AA\/ EWs exceeding 0.4\AA, and apply this discriminant to all the observations for which the strength of this feature can be measured. The adopted value generates a substantial sample of candidate cluster and tidal tail stars while minimizing the contamination from field dwarfs.  It is consistent with the results of \cite{2014MNRAS.438.3507D} who used a similar approach and a value of 0.35\AA\/ for the giant/dwarf discrimination.  Our value is also broadly consistent with the approach used in \cite{2013ApJ...764...39C}.  For those stars in our sample where the S/N of the spectrum was too low to allow a reliable measurement of the Mg I line strength, an upper limit for the EW value was adopted.

Wherever the 1700D filter was not available, the Mg~I triplet at $\sim\lambda$5180\AA\, observed in the 580V grating in the blue arm was utilized. These features can also provide gravity discrimination \citep[e.g.,][]{2013ApJ...764...39C}.  We therefore measured the total EW of the Mg I triplet lines and the resulting relation between the Mg I line strengths and the Ca II triplet EW is shown in the right panel of Fig.\ \ref{fig:ewmgca}.  Shown also in this panel are stars within the cluster radius whose Mg I $\lambda$8807\AA\/ line strengths are available.  The form of the relationship is similar to that in the left panel and we adopt a Mg I triplet EW value of 3.5\AA\/ as the value to discriminate dwarfs from giants.  The value is consistent with the dwarf/giant discrimination discussed in 
\cite{2012AJ....143...88C}, who used similar 580V observations.  

\subsubsection{Metallicity of stars}
The 2010 on-line version\footnote{http://physwww.physics.mcmaster.ca/$\sim$harris/mwgc.dat} of the Milky Way Globular Cluster catalogue \citep{1996yCat.7195....0H} lists the metallicity of Pal 5 as [Fe/H] = --1.41 dex.  This value has its origin in the Washington system photometry of \cite{1997PASP..109..799G}, which yielded [Fe/H] = --1.52 $\pm$ 0.28 (internal error), and in the high dispersion spectroscopy of 4 Pal 5 red giants analyzed by \cite{2002AJ....123.1502S} that gave [Fe/H] = --1.28 $\pm$ 0.03 (internal error).  Since there is no evidence for any metallicity dispersion in Pal 5, (e.g., \cite{2002AJ....123.1502S}) we can use metallicity as a further means to identity candidate cluster and tidal tail members.

\begin{figure*}
  \begin{center}  
    \includegraphics{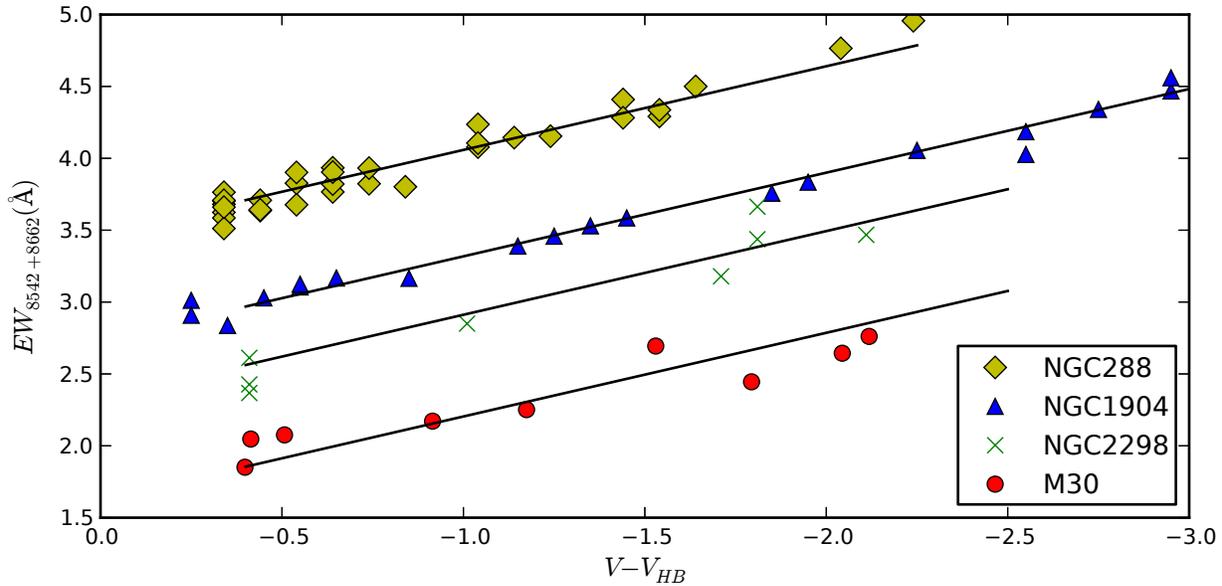}
  \end{center}
\caption{The summed EW of the $\lambda\lambda$8542 and 8662\AA\/ lines of the Ca II triplet are plotted against $V-V_{HB}$ for the calibration clusters M30 ([Fe/H] = --2.27), NGC 2298 (--1.92), NGC 1904 (--1.60) and NGC 288 (--1.32). The lines have a gradient $\alpha=-0.582$ \AA/mag. }
\label{fig:calfd}
\end{figure*}

Our metallicity determinations are based on the strength of the Ca II triplet lines in our spectra, following well-established techniques \cite[e.g.,][]{1991AJ....101.1329A}.  The calibration of the line strengths is determined from AAOmega 1700D spectra of red giants in four Galactic globular clusters obtained during other AAOmega observing programs.  The calibration clusters are, in order of increasing metallicity, M30 (NGC 7099), NGC 2298, NGC 1904 and NGC 288.  We measured the EWs of the two stronger Ca II lines in the calibration cluster spectra in the same way as for the Pal 5 program stars.  The results are shown in Fig.\ \ref{fig:calfd} in which the line strengths are plotted against $V-V_{HB}$, where $V_{HB}$ is the horizontal branch magnitude in the $V$-band for each cluster from the 2010 on-line version of the Harris (1996) catalogue.  The $V$ magnitudes of the red giant branch stars are generally taken from Stetson's on-line photometric catalogue\footnote{http://www3.cadc-ccda.hia-iha.nrc-cnrc.gc.ca/community/STETSON/standards/}. 

\begin{figure}
 \begin{center}  
    \includegraphics{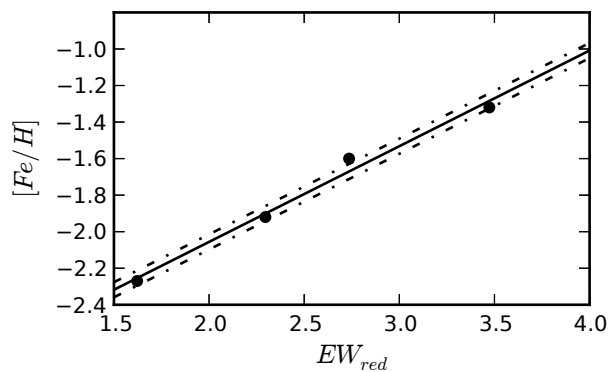}
  \end{center}
\caption{The solid circles are the reduced equivalent widths and [Fe/H] abundances for the calibration clusters.  The solid line is a linear least-squares fit to these data, while the dotted lines show the rms for the fit. }
\label{fig:calculs}
\end{figure}

The average gradient of the linear least-squares fit to the points for each calibration cluster is $\alpha=-0.58 \pm 0.03$ \AA\, mag$^{-1}$,
a value consistent with other studies. For example, \cite{2014MNRAS.441.3396Y} find $\alpha=-0.60$ from a similar set of AAOmega observations.  If we define the reduced equivalent width, $EW_{red}$ by:
\begin{equation}
EW_{red}=EW_{Ca II}+\alpha\left(V-V_{HB}\right)\label{eq:redew}
\end{equation}
where $EW_{Ca II}$ is the sum of the equivalent widths of the two stronger Ca II triplet lines, $V$ is the magnitude of the star and $V_{HB}$ is the magnitude of the horizontal branch, then the mean value of $EW_{red}$ for each calibration cluster is equivalent to the value of the relations
shown in Fig.\ \ref{fig:calfd} at $V_{HB}$ = 0.  The resulting relation between these mean $EW_{red}$ values and [Fe/H] is shown in Fig.\ 
\ref{fig:calculs}.  A linear least-squares fit to these points then yields the abundance calibration: 
\begin{equation}
[Fe/H]=(0.524 \pm 0.043)EW_{red} - (3.104\pm0.041) \label{eq:calclup5}
\end{equation}
The rms about this relation is 0.04 dex and is shown by the dotted lines in Fig.\ \ref{fig:calculs}.  Consequently, for any given program star, we can calculate $V-V_{HB}$, assuming $V_{HB}$ = 17.51 for Pal 5 \citep{1996yCat.7195....0H}, and thus the reduced equivalent width from the measured line strengths.  Equation \ref{eq:calclup5} then yields an abundance, which for members of the cluster and tidal tails, will be consistent with the known metallicity of Pal 5.  In practice, we note first that for the stars in our sample, the SDSS $ugriz$ photometry needs to be transformed to $V$ magnitudes; we use the equations given in \cite{2005AJ....130..873J}.  Second, equation \ref{eq:redew} is generally only used for stars
with $V-V_{HB}$ $<$ 0, while our sample potentially contains stars up to a magnitude fainter.  \cite{2007AJ....134.1298C}, however, have 
shown that the relation between Ca II triplet line strength and $M_{V}$ is linear to $M_{V}$ $\approx$ 1.25, i.e.,  $V-V_{HB}$ $\approx$ 0.6, 
although \cite{2012A&A...540A..27S} have suggested the relation flattens for stars fainter than $V-V_{HB}$ $\approx$ 0.3.  We have assumed the linear relation of equation \ref{eq:redew} applies for all potential $V-V_{HB}$ values.  
Third, the combination of equations \ref{eq:redew} and
\ref{eq:calclup5} strictly applies only to RGB stars.  Asymptotic giant branch (AGB) stars, however, will have weaker $EW_{Ca II}$ values at a 
given $V-V_{HB}$ because of their higher temperatures, and as a result would be assigned a lower abundance.  We have coped with these
two effects by considering as plausible for membership a range in abundance about that determined from the Pal 5 RGB members
(see \S \ref{sect3.1}).  Finally, we
note that \cite{2010A&A...513A..34S} and \cite{2007AJ....134.1298C} have shown that a linear relationship between
$EW_{red}$ and [Fe/H] is not appropriate when a large metallicity range is considered.  This is not a issue here as we are concerned only
with candidate Pal 5 members and the abundance of the cluster is within the range spanned by the calibration clusters.

We now have velocities, photometry and Ca II and Mg~I line strengths available for all the stars in our sample. In the next section, we will demonstrate how we employed these measurements to generate a list of probable members of Pal 5 and of its tidal tails. 

\section{Analysis And Discussion}  \label{sect3}
The information described in the previous section is now employed to select candidate members of the cluster and of its tidal tails.  However, the order in which the information was used was different depending on whether a particular star was a candidate cluster member or a candidate tidal tail member.  The cluster member candidates are those that fall within our adopted radius for the cluster, while the tidal tail candidates are those beyond the adopted radius, whose value we now discuss.  First, we note that using the core radius and concentration parameter given in the Harris catalogue \citep{1996yCat.7195....0H} (2010 on-line edition), the nominal tidal radius of Pal 5 is 7.6$^{\prime}$.  On the other hand, the azimuthally 
averaged surface density profile given in \cite{2003AJ....126.2385O} shows a notable change in slope at about 12$^{\prime}$ from the cluster centre, while the surface density profile in the directions perpendicular to the tidal tails show very few cluster stars beyond this radius.  \cite{2004AJ....127.2753D}, using similar data, found that the surface density profile of the cluster appeared to be truncated at $\sim$16$^{\prime}$. Furthermore, \cite{2004AJ....127.2753D} calculated a theoretical tidal radius for Pal 5 of 54 pc (8.0$^{\prime}$) at the cluster's current position, while \cite{2012A&A...546L...7M} adopted \cite{2004AJ....127.2753D}'s model A, which has tidal radius of 56 pc or 8.3$^{\prime}$ at the cluster distance of 23.3 kpc.  We decided to adopt this latter value as the radius at which to separate cluster member stars 
from stars which are likely members of the tidal tails.  None of the following analysis is strongly dependent on the actual value used for the cluster radius.

\subsection{Cluster Members} \label{sect3.1}

\begin{figure*}
  \begin{center}  
    \includegraphics{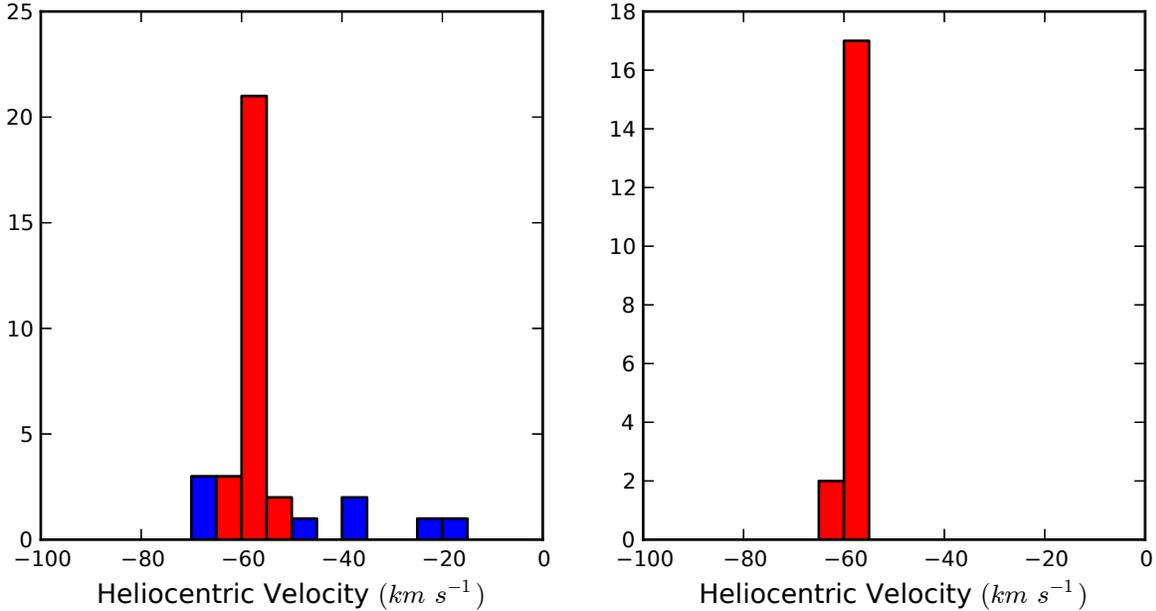}
  \end{center}
\caption{Left panel: Heliocentric velocity distribution for the red giant stars observed in fields F3 and F4 that are within a radius of 8.3$^{\prime}$ from the centre of Pal 5.  Stars that lie in the velocity range of --65 to  --50 km s$^{-1}$ are shown in red. Right panel:  The velocity distribution of the 19 stars that remain after application of the dwarf/giant and Ca II line strength selection criteria. }
\label{fig:hist}
\end{figure*}

\begin{figure}
  \begin{center}  
    \includegraphics{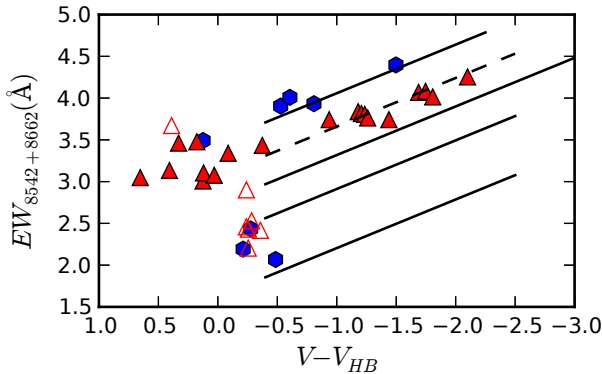}
  \end{center}
\caption{Ca II line strengths as a function of $V-V_{HB}$ are shown for the 34 stars in the left panel of Fig.\ \ref{fig:hist}.  As in that figure stars in the velocity interval --65 to --50 km s$^{-1}$ are plotted as red symbols and stars outside that velocity interval are shown as blue symbols.  The solid lines are fiducial lines for the calibration clusters from Fig.\ \ref{fig:calfd}.  The dashed line shows the fit of a line of
slope $\alpha=-0.58$ \AA/mag to the 11 probable Pal 5 member stars with $V-V_{HB} < -0.2$.  Fainter stars that are not considered likely members are shown as open red symbols.}
\label{fig:calop}
\end{figure}

In the left panel of Fig. \ref{fig:hist} we show the velocity histogram for the 34 candidate red giant stars observed in fields F3 and F4 that have velocities between --100 and zero km s$^{-1}$, and which lie within our adopted radius for the cluster.  The bin size is 5 km s$^{-1}$, which is reasonable given that our largest velocity errors are $\sim$4 km s$^{-1}$.  There is an obvious peak in the --60 to --55 km s$^{-1}$ bin, which, since \citep{2002AJ} give the velocity of Pal 5 as $-58.7 \pm 0.2$ km s$^{-1}$, is clearly dominated by Pal 5 members.  The velocity histogram, however, suggests that there remains a level of contamination from field star interlopers.  We therefore have employed our selection criteria to increase the likelihood that the stars that survive the cuts are genuine cluster members.  First, we use a photometric cut to remove stars that lie far from the principal sequences in the Pal 5 CMD\@.  We then applied our dwarf/giant separation 
criteria as illustrated in Fig.\ \ref{fig:ewmgca} to select against foreground dwarfs.  The final criteria used was the requirement for consistency between observed Ca II line strength and the known metallicity of Pal 5.  
This is illustrated in Fig.\ \ref{fig:calop}, where for completeness we show the Ca II data for all the stars in the left panel of Fig.\ \ref{fig:hist}.  As in
that figure, stars in the velocity interval --65 to  --50 km s$^{-1}$ are plotted as red symbols, while stars outside that velocity range are plotted as blue symbols.  
Inspection of Fig.\ \ref{fig:calop} shows a well-defined sequence of 11 red symbols with $V-V_{HB}$ $<$ --0.2 that lies 
between, and parallel to, the fiducial lines for the calibration clusters NGC 1904 and NGC 288.  Fitting a line of slope $\alpha$ = --0.58 \AA/mag to these data, calculating the reduced equivalent width, and applying the abundance calibration of equation \ref{eq:calclup5} then yields a mean abundance for these stars of $\langle$[Fe/H]$\rangle$ = --1.48 $\pm$ 0.10 dex. Here the uncertainty is the quadrature sum of the rms residual about the fitted line and the uncertainty in the abundance calibration.  The derived abundance agrees well with the Pal 5 abundance ([Fe/H] = --1.41) given in the
Harris catalogue and we conclude these stars are genuine Pal 5 cluster members.  For the remaining fainter stars with velocities in the --65 to  --50 km s$^{-1}$ interval, which have $V-V_{HB}$ $>$ 0, we use their location in this diagram and the errors in their line strength measures to classify eight stars as likely cluster members.  All these stars pass the CMD location and dwarf/giant separation tests, yielding a 
final sample of 19 Pal 5 cluster members.  The velocity histogram of these stars is shown in the right panel of Fig.\ \ref{fig:hist}.

\subsubsection{Cluster Kinematics} \label{sect:3.1.1}
We now use our sample of 19 probable Pal 5 red giant members to investigate the kinematics of the cluster, using a maximum likelihood technique.
Specifically, we follow the approach developed by \cite{2006AJ....131.2114W} in which the mean velocity $V_{r}$ and the intrinsic velocity
dispersion $\sigma_{cl}$ of the cluster are derived from a set of $N$ stars with velocities $\{\nu_{1},\ldots, \nu_{N}\}$ and associated errors 
$\{\sigma_{1},\ldots,\sigma_{N}\}$ via maximizing the joint probability function:
\begin{equation}
\ln{\left(p\right)}=-\frac{1}{2}\sum_{i=1}^{N}\ln{\left(\sigma_{i}^2+\sigma_{cl}^2\right)}-\frac{1}{2}\sum_{i=1}^{N}\frac{\left(\nu_{i}-V_r\right)^2}{\left(\sigma_{i}^2+\sigma_{cl}^2\right)}-\frac{N}{2} \ln{\left(2 \pi\right)} \label{eq:maxli}
\end{equation}

Application of the technique then yields a mean velocity for Pal 5 of $V_{r}=-57.4\pm0.3$ km s$^{-1}$ and a velocity dispersion $\sigma_{cl}=1.2_{-0.2}^{+0.3}$ km s$^{-1}$. The errors are determined by observing the parameter limits of the probability distribution of each variable that contains the central 68.3\%. The mean velocity has an additional uncertainty of $\pm$0.8 km s$^{-1}$ resulting from the uncertainty in
the zero point of our velocity scale (see the discussion in \cite{2012ApJ...751....6D}).  The velocity dispersion, despite its small value, represents a $>$5$\sigma$ detection. \cite{2002AJ}, using a sample of 17 members, derived a
mean velocity of --58.7 $\pm$ 0.2 km s$^{-1}$ and a velocity dispersion of 1.1 $\pm$ 0.2 km s$^{-1}$ for Pal 5.   The velocity dispersion is in excellent agreement with our determination, while there is a difference (this work -- previous) of 1.3 $\pm$ 0.4 km s$^{-1}$ in the mean velocities. Given the $\pm$0.8 km s$^{-1}$ uncertainty in the zero point of our velocity scale, this difference is not significant.

\subsubsection{Comparison with Odenkirchen et al.\  (2002)}
As noted above, \cite{2002AJ} analysed the kinematics of Pal 5 using radial velocities derived from spectra taken with the UVES instrument on
the VLT\@.  From the 20 stars observed, all of which lie within 6$^{\prime}$ of the cluster centre, 17 were classified as Pal 5 members
primarily on the basis of radial velocity.  Our Pal 5 sample includes 13 of these 17 stars, the other four were not included in the 2dF configurations.  We categorize 12 of the stars in common as members of Pal 5, while one star, star 12 in O02 sample, we classify as non-member on the basis of a Mg I line strength that exceeds our threshold.  Our velocities and those of O02 are given in Table \ref{tab:resioden02} for the stars in common, together with our membership classification.  The (unweighted) mean velocity difference, in the sense (this work -- O02), is 0.9 $\pm$ 0.4 km s$^{-1}$ (std error of the mean), which is consistent with the difference in mean cluster velocity derived above.  None of the velocity differences exceeds three times the standard deviation.  The properties of the full sample of 19 Pal 5 red giant members, 12 in common with O02 and 7 new Pal 5 stars discovered in our analysis, are given in the first part of Table \ref{tab:targetstable}.  The velocities presented are those derived here.

\begin{table*}

\begin{minipage}{160mm}
\begin{center}
\caption{A comparison of the radial velocities derived here with those given by \citet{2002AJ} for the Pal 5 stars in common. }

\label{tab:resioden02}
\begin{tabular}{@{}ccccccc}

\hline \hline
Target & Star & $\nu_{helio}$ & $\sigma$ & $\nu_{helio}\,$ & $\sigma$&Member$^{a}$\\
& O02 & (km s$^{-1}$) &(km s$^{-1}$) & O02 (km s$^{-1}$)$^{a}$ &(km s$^{-1}$) & \\
\hline
P1205646&1&-57.9&0.7&--58.51&0.05&Yes\\
P1201565&2&-56.8&0.7&--58.31&0.05&Yes\\
P1206550&5&-56.0&0.6&--56.92&0.07&Yes\\
Pal5\_229p5Oden148&6&--59.0&0.4&-58.72&0.09&Yes\\
P1203635&7&-57.7&0.8&--58.79&0.10&Yes\\
P1207360&9&-56.9&0.4&--57.35&0.15&Yes\\
P1204797&10&-56.4&1.0&--60.10&0.14&Yes\\
P1205893&11&-57.6&1.6&--58.90&0.09&Yes\\
Pal5\_229p5Oden140&12&--55.6&3.9&-52.97&0.12&No\\
P1201384&13&-57.7&0.3&--58.98&0.08&Yes\\
Pal5\_229p5Oden153&14&--60.9&3.3&-61.07&0.17&Yes\\
P1202728&17&-57.2&2.7&--58.54&0.06&Yes\\
P1205197&20&-55.6&1.5&--57.27&0.06&Yes\\
\hline

\end{tabular}
\end{center}
$^{a}$ Membership status as determined by applying our selection criteria.

\end{minipage}
\end{table*}

\subsection{Tidal Tail Members}

Although there is some existing information on the kinematics of the stars in the Pal 5 tidal tails \citep{2009AJ....137.3378O}, our spatial 
coverage is considerably more extensive.  Consequently, it is not appropriate to use radial velocity as the primary selection criterion for membership in the tidal tails, rather we use the other selection criteria (Giant/Dwarf separation, CMD location and Ca II line strengths) first, and then investigate the radial velocities of the remaining stars.  Specifically, we first excluded stars whose Mg~I line strengths were above the cutoffs shown in Fig.\ \ref{fig:ewmgca}.  Second, we required consistency with the location of the Pal 5 member stars in the (Ca II line strength, $V-V_{HB}$) diagram (Fig.\ \ref{fig:calop}).  
Then we required consistency between the colour and magnitude of the stars and the principal sequences in the CMD of the cluster.  Both these latter criteria implicitly assume that there is no substantial variation in distance from the Sun along the tidal tails compared to the distance to the cluster. This
assumption is consistent with the results of \cite{2006ApJ...641L..37G}, who, on the basis of their best-fit orbital model, indicate variations in the tidal tail distance modulus relative to that of the cluster of order 0.06 mag or less.  Similarly, model A of \cite{2004AJ....127.2753D} shows variations in the distance modulus of the tidal tails that are relatively minor, insufficient to significantly affect the selection process.  Only after we have a set of stars meeting these criteria did we consider the radial velocities, requiring candidate tidal tail members to have velocities in the range $-70$ to $-35$ km s$^{-1}$.  This process resulted in a final sample of 47 candidate tidal tail members, 30 stars in the trailing tail and 17 stars in the leading tail.  These candidate tidal tail stars cover the full extent of the area surveyed: for example, the most distant trailing tail star is $\sim$17.5 deg, or 7.1 kpc in projection, from the centre of Pal 5. As an added check, we matched our candidates with the PPMXL \citep{2010AJ....139.2440R} and UCAC4 \citep{2013AJ....145...44Z} catalogues. None of our stars possess measured proper motions in excess of the errors, consistent with their classification as (distant) giants.

In each of the 2dF fields in the outer parts of the trailing tail we have typically identified 3 candidate members per field.  Given these small numbers we have used the Besan\c{c}on model of the Galaxy \citep{2003A&A...409..523R} to investigate the extent to which any of our candidates might actually be field stars that, by chance, happen to meet our selection criteria.  We generated 10 realizations from the Besan\c{c}on model using the location of the outermost field F11 and the 2dF field-of-view.  For each of these models we then randomly selected 10 sets of 330 stars (i.e., a typical observed sample) that lie within the selection window shown in the left panel of Fig. 1.  Each set was then evaluated and the number of giants (i.e., log g < 3) with metallicities and velocities within our adopted ranges for Pal 5 recorded.  We found that of the 100 trials, there were 11 occurrences where one model star met all our criteria and none where two or three were selected.  Consequently, while we cannot rule out a minor level of contamination of our outer sample from field stars, it is not sufficient to significantly bias the results.

 \subsubsection{Comparison with Odenkirchen et al.\ 2009} \label{sect:O09comp}
In their study of the kinematics of the Pal 5 tidal tails, O09 identified 17 likely leading and trailing tidal tail members in a region covering approximately 8.5 deg on the sky.  A comparison with our list of observed stars revealed 11 stars in common with O09, although 3 were subsequently discarded from our analysis as their spectra had unacceptably low signal-to-noise ratios.  Reassuringly, the remaining 8 O09 stars were also classified as tidal tail members in our analysis.  Table \ref{tab:resien} compares our velocity measurements with the O09 values.   Including all 8 stars, the
(unweighted) mean velocity difference, in the sense (this work -- O09), is 1.5 $\pm$ 0.9 km s$^{-1}$ (std error of the mean).  This value is 
consistent with that for the cluster-star comparison.  The largest difference is for O09 star 30218 for which we find a velocity of --53.8 $\pm$ 1.9 km s$^{-1}$, 7.1 km s$^{-1}$ higher than the --60.9 $\pm$ 0.3 km s$^{-1}$ velocity given by O09.  This star may be a binary.  If it is excluded
from the comparison, the mean velocity difference becomes 0.7 $\pm$ 0.6 km s$^{-1}$, indicating excellent agreement.

\begin{table*}

\begin{minipage}{160mm}
\begin{center}
\caption{A comparison of the radial velocities derived here with those give by \citet{2009AJ....137.3378O} for the Pal 5 tidal tail stars in common.}

\label{tab:resien}
\begin{tabular}{@{}ccccccc}

\hline \hline
Target & Star & $\nu_{helio}$ & $\sigma$ & $\nu_{helio}$ & $\sigma$&Member$^{a}$\\
& O09 & (km s$^{-1}$)& (km s$^{-1}$) & O02 (km s$^{-1}$)$^{a}$ & (km s$^{-1}$) & \\

\hline
P1203859&20015&-58.7&1.4&--58.20&0.47&Yes\\
P1227758&20017&-59.1&1.3&--57.82&0.31&Yes\\
P1231315&20016&-56.7&2.3&--58.27&0.48&Yes\\
Pal5\_231Oden\_7&31076&--50.0&2.4&-51.73&0.22&Yes\\
Pal5\_231Oden\_6&20006&--56.5&2.6&-56.70&0.18&Yes\\
Pal5\_231Oden\_2&31023&--55.6&0.4&-55.44&0.22&Yes\\
Pal5S\_113&30076&-57.8&1.1&--60.72&0.35&Yes\\
Pal5S\_210&30218&-53.8&1.9&--60.89&0.28&Yes\\ 
\hline

\end{tabular}

\end{center}
$^{a}$ Membership status as determined by applying our selection criteria.
\end{minipage}
\end{table*}

\subsection{Blue Horizontal Branch Stars}
As is apparent from the selection box in the left panel of Fig.\ \ref{fig:p5cmd}, a number of the fields observed with 2dF include stars that are potentially blue horizontal branch (BHB) members of the cluster and of the tidal tails.  As these stars are hotter than their red giant counterparts, a different analysis approach was required.   Clearly the requirement for agreement in colour and magnitude with the Pal 5 CMD sequence remains valid, but no metallicity estimate is possible as the Ca II triplet spectral region is now dominated by the hydrogen lines from the Paschen series.  The principal contaminant of the candidate BHB stars are foreground blue straggler stars.  These were distinguished from genuine BHB candidates through the characteristically broader hydrogen lines of the higher gravity stars.  Radial velocities were again calculated by cross-correlation, but this time the template employed was a high S/N
spectrum of the field BHB star HD 86986, observed with the 1700D grating setup as part of a separate AAOmega program.  We adopted the 
radial velocity given by SIMBAD for the star (9.5 $\pm$ 0.4 km s$^{-1}$); the uncertainty in the zero point of the resulting velocities is unlikely to exceed 2--3 km s$^{-1}$.  Candidate BHB members of the cluster and the tidal tails were then chosen (after applying the photometric and gravity selection) on the basis of having velocities similar to velocities of red giant candidates in the same 2dF field.  In the end only one BHB candidate was identified through this process.  It lies just within our adopted cluster radius and its properties are given in the first part of Table \ref{tab:targetstable}.  We have chosen not  to include this star in the discussion of the cluster kinematics (see \S \ref{sect:3.1.1}) because of the uncertainty in whether the velocity is on the same system as that for the red giants.

\subsection{Final Tidal Tail Sample}
\begin{figure}
  \begin{center}  
    \includegraphics[]{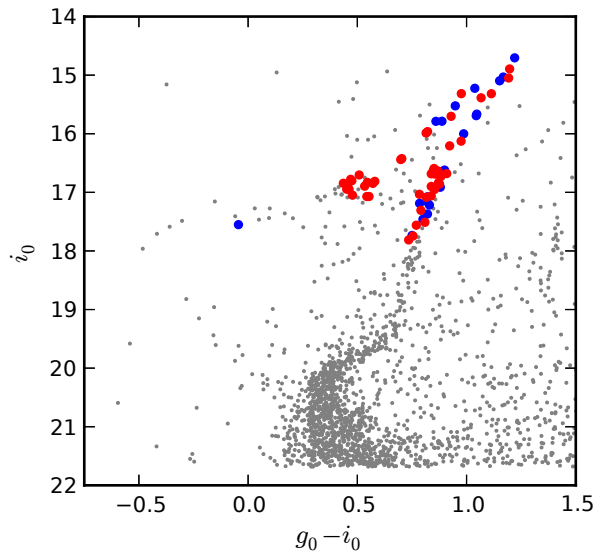}
  \end{center}
\caption{Location of candidate cluster and tidal tail member stars in the Pal 5 CMD from Fig.\ \ref{fig:p5cmd}.  Blue points indicate cluster stars, i.e., those within 8.3$^{\prime}$ of the cluster centre.  The red points show the tidal tail stars.}
\label{fig:seleected}
\end{figure}

\begin{figure*}
  \begin{center} 
    \includegraphics{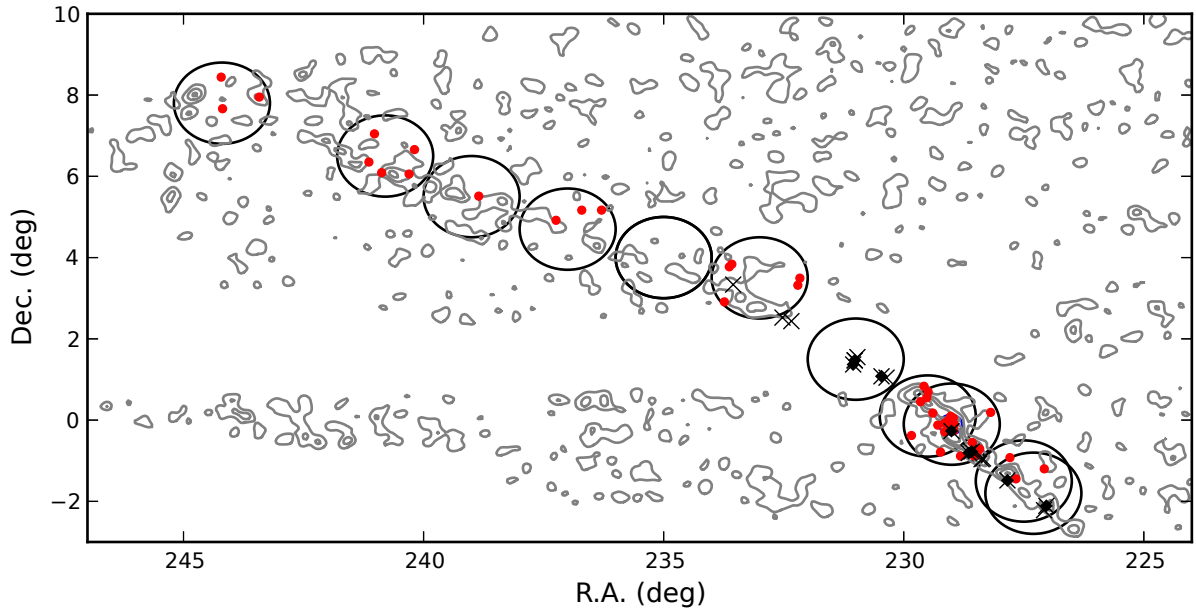}
  \end{center}
\caption{The positions of candidate members of the cluster, and of the tidal tails, compared to the surface density contours from \citet{2006ApJ...641L..37G}.  Red points indicate candidates found in this work.  Black crosses show the tidal tail candidate stars from O09 and a superposed black dot indicates the star were also found in this study.  The large circles indicate the 2dF fields studied, which are designated F1 -- F11 in order of increasing Right Ascension. }

\label{fig:contour}
\end{figure*}

The properties of the full sample of 47 Pal 5 tidal tail candidate members, of which 39 are new, are given in the second part of Table \ref{tab:targetstable}, where, for convenience, we present the stars in the leading and trailing tails separately.  The location
of these stars, and the cluster member candidates, are shown in the reddening corrected Pal 5 CMD of Fig.\ \ref{fig:seleected}.  Not surprisingly, the tidal tail stars conform closely to the principal cluster sequences in the CMD\@.  In Fig.\ \ref{fig:contour} we show the positions of the Pal 5 cluster and tidal tail stars
over plotted on the contour diagram of \citet{2006ApJ...641L..37G}, which uses photometry from SDSS DR4.  Within approximately one degree or so of the cluster centre, our candidates align well with the surface density contours, which lie well above the background.  At larger distances from the cluster, however, there is less of a correspondence between the location of the individual stars and the density contours, which lie closer to the background density.  This comparative lack of correspondence can be ascribed to the small number of our candidates.  A larger sample of fainter spectroscopically confirmed tidal tail stars should nevertheless coincide with the density contours, which, in the analyis of \citet{2006ApJ...641L..37G}, are dominated by the location of the much numerous, as compared to red giants, main sequence stars.

\subsection{Velocity Gradient and Dispersion}

In their study of the kinematics of the Pal 5 tidal tails, O09 revealed the presence of a linear gradient in the line-of-sight velocities of the tidal tail stars with angular position along the stream -- velocities of the leading tail stars were more negative than that of the cluster, which in turn was more negative than the velocities of the trailing tail stars.  The gradient determined was $1.0 \pm 0.4$ km s$^{-1}$ deg$^{-1}$ across an arc 
approximately 8.5 deg in extent, and for a sample of 15 candidate tidal tail members.  O09 also determined the velocity dispersion in the tidal
tails finding a value of 2.2 km s$^{-1}$ for the same sample of 15 candidates, demonstrating that the tidal tails are a kinematically cold structure.
Our study of the tidal tails covers a much larger angular extent than that of O09, particularly as regards the trailing tail, and thus it is important to evaluate whether the velocity gradient and the low velocity dispersion persist with increasing distance from the cluster centre.

\begin{figure}
  \begin{center}  
    \includegraphics{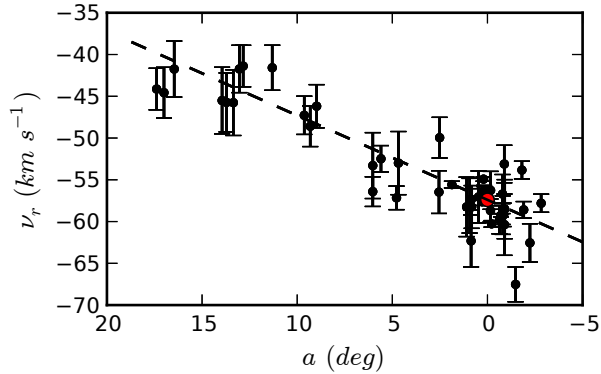}
  \end{center}
\caption{The radial velocities of the 47 Pal 5 tidal tail stars, i.e., those further than 8.3$^{\prime}$ from the cluster centre, are plotted against {\it a}, the angular distance from the cluster centre in degrees.  The red dot represents the mean velocity of the cluster stars, while the dashed line is the derived overall velocity gradient of  $1.0\pm0.1$ km s$^{-1}$deg$^{-1}$.}
\label{fig:velgrad}
\end{figure}

In Fig.\ \ref{fig:velgrad} we show the velocities of our tidal tail stars, i.e., those outside the cluster radius of 8.3$^{\prime}$, against {\it a}, which we use to denote the angular distance in degrees of each star from the centre of Pal 5.  We note that through the choice of the 2dF field centres, the candidate stars are at most      
$\sim$1 deg from the nominal stream centre in their vicinity, so that we can use {\it a} as valid measure of angular distance along the tidal tails.  We note also that {\it a} is positive in the trailing tail and negative in the leading tail, and that it is essentially equivalent to the quantity $\Delta l \cos{b}$ used by O09 for the region
of the tails covered in their analysis.  

If we consider first the region $-3^{\circ}< a < 6.5^{\circ}$, which coincides with the section of the tidal tails covered by O09, we find for our sample of 35 stars a linear gradient between $\nu_{r}$ and {\it a} of $0.9 \pm 0.3$ km s$^{-1}$ deg$^{-1}$ through a weighted least-squares fit, where the weights are the inverse square of the velocity errors.  This value is quite consistent with the gradient, $1.0 \pm 0.4$ km s$^{-1}$ deg$^{-1}$ found
by O09.  If we add to the sample the 7 tidal tail stars in O09's sample of 15 not observed by us, after adjusting their velocities by 0.7 km s$^{-1}$ (see \S \ref{sect:O09comp}), the derived gradient is only marginally different $0.9 \pm 0.1$ km s$^{-1}$ deg$^{-1}$. As regards the intrinsic velocity dispersion about this velocity gradient, we used the maximum likelihood approach described in \S \ref{sect:3.1.1} after first correcting the observed velocities by the velocity predicted at the {\it a} value of each star by linear velocity gradient.  Such an approach is necessary since our velocity errors are notably larger than those of O09.  We find that the intrinsic velocity dispersion in this region of the tidal arms is $2.0 \pm 0.4$ km s$^{-1}$, entirely consistent with the value of 2.2 km s$^{-1}$ (no error given) found by O09.  Increasing the sample with the additional 7 O09 stars does appreciably change this value.  

We now consider the 12 trailing tail stars that lie beyond {\it a} $\approx$ 6 deg, noting that currently there are no stars known in the leading tail at these distances from the cluster centre.  Including the mean velocity for the cluster at {\it a} = 0, the derived gradient is $1.1 \pm 0.1$ km s$^{-1}$ deg$^{-1}$ which is not significantly different from the gradient shown by the inner sections of the tail. The intrinsic velocity dispersion was undetectable given the uncertainties of velocities. However, we can limit the velocity dispersion to less than 4.2 km s$^{-1}$, still characteristically low.  If we consider only the data for these stars without including the mean velocity at {\it a} = 0, the calculated gradient becomes $0.4 \pm 0.2$ km s$^{-1}$ deg$^{-1}$.  This might indicate a decrease in the size of the velocity gradient in the outer parts of the trailing tail.

We now turn to determining the gradient and dispersion over the full almost 20 deg arc of tidal tails using the full sample of 47 tidal tail stars observed here.  A first order fit yields a gradient of $1.0 \pm 0.1$ km s$^{-1}$ deg$^{-1}$ with an intrinsic dispersion about the gradient of $2.1 \pm 0.4$ km s$^{-1}$.  The fit is shown in Fig. \ref{fig:velgrad}.  Again these values are not appreciably different from those given in O09 despite the larger sample and the larger angular coverage.  We also considered a quadratic fit to the data, but found that the quadratic term was not significant.

Our results reinforce the identification of the Pal 5 tidal tails as a kinematically cold structure, at least for the sections of the tidal tails that have kinematic data.  The recent results of \cite{2012ApJ...760...75C} show that the trailing tail continues as a narrow feature for a total length of at least $\sim$23 deg; it may extend even longer as the \cite{2012ApJ...760...75C} analysis is limited by the boundary of the SDSS survey region.  Kinematic studies of this extended region will be difficult, however, as the contrast of the tail features compared to the background is much reduced at lower Galactic latitudes.  The current data on the surface density of the leading tail is also limited, at an angular extent of $\sim$6 deg, by the boundary of the SDSS survey region.  In this case it is because the SDSS survey region does not penetrate to any significant extent south of the equator.   The SkyMapper survey of the southern hemisphere sky \citep{2007PASA...24....1K} will, however, allow the leading tail to be mapped into the southern hemisphere.  The SkyMapper filter system is designed to provide gravity and metallicity information \citep{2007PASA...24....1K} for survey stars, which should facilitate the selection of candidates for spectroscopic follow-up.  It will be intriguing to see if a single velocity gradient and a constant velocity dispersion remain the best interpretation of the data when kinematic information is available for a comparable angular distance in the leading tail as is currently available for the trailing tail.  Together with the results presented in this paper, such additional data would provide strong constraints on the orbit of Pal 5, on the tidal disruption process, and on the Galaxy's dark matter halo.

\section{Conclusion}

We have demonstrated here a detailed method for identifying members of Pal 5 and of its tidal tails.  The approach distinguishes candidate members from contaminating field stars through a combination of kinematic, line strength and photometric information.  The result is the selection of 67 candidate members of the cluster and its tidal tails, of which 47 are new objects.  The sample consists of seven new red giant and one new BHB  members lying within 8.3$^{\prime}$ of the cluster centre, twelve reconfirmed cluster members, 27 new members of the trailing tail, three reconfirmed trailing tail stars, and 12 members of leading tail of which five were previously known. Our overall coverage is $\sim$20 deg along the tails, with the coverage of the trailing tail being substantially larger than in previous work.

For the Pal 5 cluster members we derive, through a maximum likelihood technique, a mean velocity of $-57.4 \pm 0.3$ km s$^{-1}$ and an intrinsic velocity dispersion of 1.2 $\pm$ 0.3 km s$^{-1}$, values that are consistent with previous determinations.  Within the region of the tidal tails studied by O09, we find the same velocity gradient and velocity dispersion.  Our angular coverage of the tidal tail is, however, considerably larger yet intriguingly we find that the velocity gradient and velocity dispersion do not change significantly from the O09 values.  Our determination is a linear velocity gradient of 1.0 $\pm$ 0.1 km s$^{-1}$ deg$^{-1}$ and an intrinsic dispersion about this gradient of $2.1 \pm 0.4$ km s$^{-1}$ across the almost 20 deg arc of the tidal tails studied here, although there is some indication that the gradient may be less at larger angular distances.  The Pal 5 tidal tails are indeed kinematically cold structures.  We note, however, that coverage of the leading tail is much less than that of the trailing tail, and we look forward to the outcomes of southern hemisphere sky surveys such as SkyMapper that will redress the situation.  In summary, the results presented here provide a promising opportunity to further constrain the tidal disruption process, the orbit of Pal 5 and of the tidal tail stars, and in particular, the properties of the Galactic halo.

\begin{table*}
\begin{minipage}{160mm}
\begin{center}
\caption{A list of all stars determined to be members in this work.}
\label{tab:targetstable}
\begin{tabular}{cccccccc}
\multicolumn{8}{ c }{Cluster Members} \\
\hline \hline
Star Designation&R.A.&Dec&$V_{r}$&$\sigma$&Mag$^{a}$&Mag$^{a}$&New Member$^{b}$\\
& (Deg, J2000) & (Deg, J2000) &(km s$^{-1}$) & (km s$^{-1}$) &(g$-$i)& (i) & \\
\hline
P1207696&15 15 48.19&--00 06 07&--55.9&1.2&0.84&17.85&Yes\\
P1207360&15 15 49.70&--00 07 01&--56.9&0.4&0.95&15.90&O02\\
P1206550&15 15 52.60&--00 07 40&--56.0&0.6&1.04&15.64&O02\\
P1205893&15 15 54.79&--00 06 55&--57.6&1.6&0.99&16.73&O02\\
P1205646&15 15 56.11&--00 06 06&--57.9&0.7&1.26&15.15&O02\\
P1205197&15 15 57.05&--00 08 50&--55.6&1.5&0.89&17.58&O02\\
P1204797&15 15 58.26&--00 09 47&--56.4&1.0&0.97&17.03&O02\\
P1204574&15 15 58.89&--00 05 17&--55.9&0.3&1.31&14.82&Yes\\
Pal5\_229p5Oden148&15 15 59.52&--00 09 00&--59.0&0.4&1.08&16.12&O02\\
P1203635&15 16 02.00&--00 08 03&--58.0&0.8&0.97&16.75&O02\\
P1203153&15 16 03.61&--00 07 17&--57.7&2.5&1.14&15.81&Yes\\
P1202728&15 16 04.81&--00 06 28&--57.2&2.7&0.93&17.33&O02\\
P1202285&15 16 06.54&--00 07 01&--57.1&0.2&1.07&15.43&Yes\\
P1202039&15 16 07.75&--00 10 18&--60.6&0.3&1.14&15.78&Yes\\
Pal5\_229p5Oden153&15 16 08.51&--00 05 10&--60.9&3.3&1.13&15.34&O02\\
P1198266&15 16 19.83&--00 01 08&--56.9&1.0&0.98&15.90&Yes\\
P1197361&15 16 23.11&--00 03 24&--58.0&2.5&0.88&17.31&Yes\\
P1194244&15 16 34.71&--00 04 25&--58.2&3.8&0.92&17.20&Yes\\
P1201565&15 16 08.66&--00 08 03&--56.8&0.7&0.91&17.27&O02\\
P1201384&15 16 09.58&--00 02 40&--57.7&0.3&1.25&15.21&O02\\
\hline
\\
\multicolumn{8}{ c }{Trailing tail Members} \\
\hline \hline
P1206410&15 15 52.84&00 03 22&--55.7&0.6&1.20&15.43&Yes\\
P1205450&15 15 56.21&00 02 25&--55.9&0.7&0.94&17.06&Yes\\
P1202648&15 16 05.26&00 05 42&--54.3&0.3&0.92&17.49&Yes\\
P1184727&15 17 09.99&--00 07 25&--57.0&0.4&0.93&16.10&Yes\\
P1178230&15 17 34.55&00 10 26&--56.5&3.1&1.02&16.81&Yes\\
P1172002&15 17 58.32&00 41 56&--57.5&3.3&0.86&17.67&Yes\\
Pal5\_229p5\_224&15 18 04.96&00 33 05&--57.7&2.6&1.04&16.43&Yes\\
P1166550&15 18 18.92&00 49 58&--57.6&3.6&0.97&16.84&Yes\\
P1161723&15 18 35.89&00 27 10&--56.7&1.3&1.30&15.01&Yes\\
P1149198&15 19 21.42&--00 22 43&--61.7&3.1&0.65&17.20&Yes\\
Pal5\_231Oden\_2&15 21 51.16&01 04 43&--55.6&0.4&0.94&16.79&O09\\
Pal5\_231Oden\_6&15 24 04.85&01 28 13&--56.5&2.6&0.96&16.78&O09\\
Pal5\_231Oden\_7&15 24 13.00&01 22 09&--50.0&2.4&0.96&16.72&O09\\
P1001405&15 28 39.20&03 29 37&--56.6&1.4&0.93&17.18&Yes\\
P0997712&15 28 49.34&03 19 10&--52.4&3.8&0.64&17.18&Yes\\
P0901878&15 34 19.31&03 50 15&--55.8&1.8&0.92&16.78&Yes\\
P0897761&15 34 31.90&03 46 24&--52.7&4.0&0.63&16.93&Yes\\
P0885066&15 34 56.51&02 54 34&--51.9&1.6&1.28&15.14&Yes\\
P0706090&15 45 10.57&05 10 06&--45.6&2.6&0.57&17.16&Yes\\
P0672715&15 46 49.44&05 10 01&--48.0&2.4&0.56&17.07&Yes\\
P0626179&15 48 57.99&04 55 04&--46.7&2.3&0.69&17.00&Yes\\
P0503947&15 55 24.13&05 30 47&--41.0&2.7&0.55&16.98&Yes\\
P0420925&16 00 45.41&06 39 29&--41.1&2.8&0.79&16.55&Yes\\
P0404276&16 01 12.59&06 03 23&--40.8&2.5&1.02&15.81&Yes\\
P0364040&16 03 29.59&06 05 26&--45.2&3.9&0.93&17.01&Yes\\
P0366470&16 04 05.53&07 02 43&--44.9&4.0&0.54&17.06&Yes\\
P0347248&16 04 33.28&06 21 07&--45.1&3.5&0.57&16.91&Yes\\
P0226451&16 13 40.97&07 57 05&--41.1&3.4&0.53&16.96&Yes\\
P0172773&16 16 44.79&07 39 50&--44.0&3.0&0.68&16.93&Yes\\
P0185610&16 16 51.73&08 26 29&--43.5&2.5&0.96&16.94&Yes\\
\hline
\end{tabular}
\end{center}
$^{a}$ From SDSS catalogue.\\
$^{b}$ Denotes the origin of the membership classification: Yes for this paper, otherwise O02 or O09.
\end{minipage}

\end{table*}

\begin{table*}
\centering
\begin{minipage}{160mm}
\begin{center}
\contcaption{}
\label{tab:resioden}
\begin{tabular}{cccccccc}
\multicolumn{8}{ c }{Leading tail Members} \\
\hline \hline
Star Designation&R.A.&Dec&$V_{r}$&$\sigma$&Mag$^{a}$&Mag$^{a}$&New Member$^{b}$\\
& (Deg, J2000) & (Deg, J2000) &(km s$^{-1}$) & (km s$^{-1}$) &($g-i$)& ($i$) & \\
\hline
Pal5S\_113&15 08 07.15&--02 06 39&--57.8&1.1&1.00&16.82&O09\\
P1340251&15 08 17.50&--01 11 59&--62.0&2.3&0.63&16.85&Yes\\
P1298414&15 10 39.02&--01 26 45&--58.0&1.0&1.11&16.29&Yes\\
P1288424&15 11 09.04&--00 55 24&--66.9&2.1&0.92&16.11&Yes\\
Pal5S\_210&15 11 21.70&--01 29 02&--53.8&1.1&1.03&17.04&O09\\
P1258991&15 12 45.44&00 11 23&--52.5&2.2&0.56&16.89&Yes\\
P1242532&15 13 40.44&--00 42 29&--58.6&1.4&1.04&16.35&Yes\\
P1238216&15 13 54.40&--00 49 09&--59.8&3.7&0.85&17.95&Yes\\
P1234307&15 14 09.32&--00 52 40&--57.9&3.5&0.87&17.89&Yes\\
P1232006&15 14 17.18&--00 33 08&--59.2&1.7&0.92&17.65&Yes\\
P1231315&15 14 20.71&--00 46 22&--56.7&2.3&0.94&17.23&O09\\
P1227758&15 14 34.63&--00 48 27&--59.1&1.3&0.91&17.46&O09\\
P1216792&15 15 16.47&--00 53 10&--58.6&2.3&0.65&17.04&Yes\\
P1207226&15 15 50.43&--00 15 45&--55.7&2.3&0.88&17.15&Yes\\
P1203859&15 16 01.54&--00 16 08&--58.7&1.4&1.08&16.13&O09\\
P1194127&15 16 34.95&--00 17 26&--59.7&0.4&1.17&15.51&Yes\\
P1188878&15 16 56.20&--00 47 20&--58.7&0.2&0.82&16.57&Yes\\
\hline

\end{tabular}
\end{center}
$^{a}$ From SDSS catalogue.\\
$^{b}$ Denotes the origin of the membership classification: Yes for this paper, otherwise O02 or O09.
\end{minipage}
\end{table*}

\section{Acknowledgments}
This research has been supported in part by the Australian Research Council through Discovery Projects grant DP120101237.
We thank Dr.\ Carl Grillmair for providing a digital copy of the density contour diagram shown in Fig.\ \ref{fig:contour}. This research has made use of the VizieR catalogue access tool, CDS, Strasbourg, France.

Funding for SDSS-III has been provided by the Alfred P. Sloan Foundation, the Participating Institutions, the National Science Foundation, and the U.S. Department of Energy Office of Science. The SDSS-III web site is http://www.sdss3.org/.

SDSS-III is managed by the Astrophysical Research Consortium for the Participating Institutions of the SDSS-III Collaboration including the University of Arizona, the Brazilian Participation Group, Brookhaven National Laboratory, Carnegie Mellon University, University of Florida, the French Participation Group, the German Participation Group, Harvard University, the Instituto de Astrofisica de Canarias, the Michigan State/Notre Dame/JINA Participation Group, Johns Hopkins University, Lawrence Berkeley National Laboratory, Max Planck Institute for Astrophysics, Max Planck Institute for Extraterrestrial Physics, New Mexico State University, New York University, Ohio State University, Pennsylvania State University, University of Portsmouth, Princeton University, the Spanish Participation Group, University of Tokyo, University of Utah, Vanderbilt University, University of Virginia, University of Washington, and Yale University.

\bibliographystyle{mn2e-2}
\bibliography{mybibs}

\label{lastpage}

\end{document}